\documentclass[publish,JDR,paper]{jdrarticle}
\usepackage{graphicx}
\usepackage{epsfig}
\usepackage{url}
\usepackage{cite}
\usepackage{amsmath,amssymb,amsfonts}
\usepackage{algorithmic}

\SetVolumeNumber{Vol.0 No.0, 200x} 
\SetArticleID{Dr*-*-****; \jtoday}

\begin{document}

\pagestyle{jdrstyle}

\title{Effectiveness of the COVID-19 Contact-Confirming Application (COCOA) Based on a Multi-Agent Simulation}

\author{Yuto Omae$^{*1}$, Jun Toyotani$^{*1}$, Kazuyuki Hara$^{*1}$, Yasuhiro Gon$^{*2}$, Hirotaka Takahashi$^{*3}$}
\address{

$^{*1}$
College of Industrial Technology, Nihon University, 1-2-1, Izumi, Narashino, Chiba, 275-8575 Japan\\

$^{*2}$
Nihon University School of Medicine, 30-1, Kami, Ooyaguchi, Itabashi, Tokyo, 173-8610 Japan\\

$^{*3}$
Graduate School of Integrative Science and Engineering, Tokyo City University, 1-28-1, Tamazutsumi, Setagaya, Tokyo 158-0087 Japan

         E-mail: oomae.yuuto@nihon-u.ac.jp}
\markboth{Y. Omae}{Effectiveness of the COVID-19 Contact-Confirming Application (COCOA) Based on a Multi-Agent Simulation}
\dates{00/00/00}{00/00/00}


\maketitle

\begin{abstract}
\noindent Abstract. 
As of Aug. 2020, coronavirus disease 2019 (COVID-19) is still spreading in the world.
In Japan, the Ministry of Health, Labor, and Welfare developed ``COVID-19 Contact-Confirming Application (COCOA),'' which was released on Jun. 19, 2020.
By utilizing COCOA, users can know whether or not they had contact with infected persons.
If those who had contact with infectors keep staying at home, they may not infect those outside.
However, effectiveness decreasing the number of infectors depending on the app's various usage parameters is not clear.
If it is clear, we could set the objective value of the app's usage parameters (e.g., the usage rate of the total populations) and call for installation of the app. 
Therefore, we develop a multi-agent simulator that can express COVID-19 spreading and usage of the apps, such as COCOA.
In this study, we describe the simulator and the effectiveness of the app in various scenarios.
The result obtained in this study supports those of previously conducted studies.
\end{abstract}

\vspace{-10pt}
\begin{keywords}
COVID-19, Contact-Confirming Application, Multi-Agent Simulation, SEIR Model
\end{keywords}

\vspace{-10pt}
\section{Introduction}
As of Aug. 2020, coronavirus disease 2019 (COVID-19) infection is still spreading in the world.
To overcome the spread of COVID-19, the Japanese government gave stay at home order to its citizens from Apr. 7, 2020 to May. 25, 2020\cite{ref:kinkyu}.
Consequently, the number of total infectors significantly decreased
(The number of daily average infectors one week ago before the order was about 300 persons; moreover, the number of daily average infector one week later after issuing the order was about 40 persons\cite{opdat}.).
Besides, as measures to overcome the spreading of COVID-19, the Japanese government is distributing cloth masks, as well as supporting the sterilization of medical institutions and PCR testing\cite{ref:taisaku}.

\begin{figure}[b]
\centering
\includegraphics[scale=0.45]{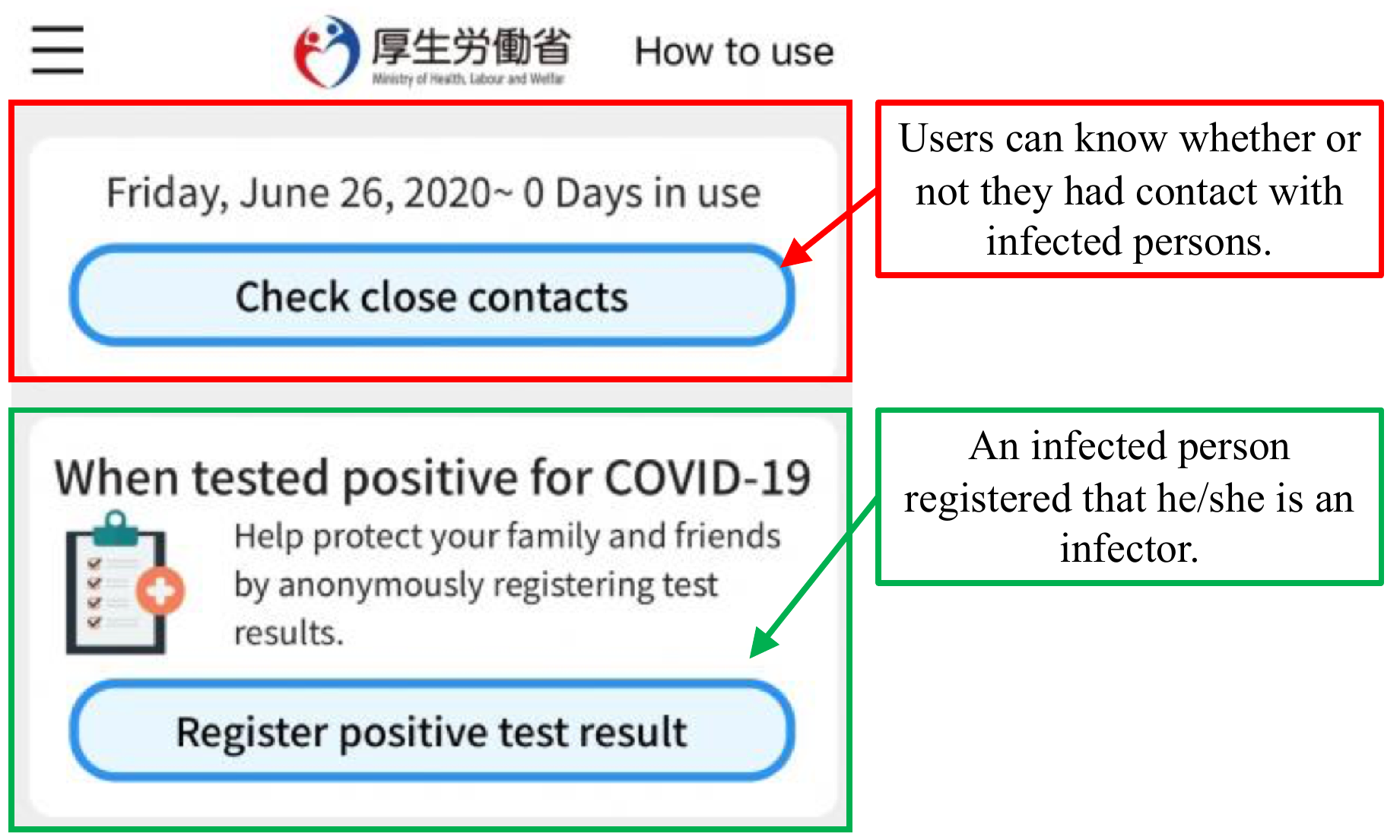}
\caption{The display of the app (COCOA).}
\label{fig:disp}
\end{figure}

In addition, the Japanese government developed ``COVID-19 Contact-Confirming Application (COCOA).''
COCOA is an application that can be installed into a smartphone, and by utilizing it, users can know whether they had contact with infectors of COVID-19.
Hereinafter, this is called an app.
The diagram of the app is shown in {\bf Fig.\ref{fig:disp}}.
The upside of the figure is used to verify whether the user had contact with the infectors of COVID-19.
If the user is infected with COVID-19, he/she registers via the button shown in the downside of the figure.
If infectors register, other users who are not infected can know whether or not to have contact with infectors for at least two weeks or less via the app.
To obtain this information, one is required to install the app and turn on the bluetooth of his/her smartphone.
Moreover, it is necessary that infectors and other persons approaching within 1 meter and over 15 minutes \cite{ref:COCOA_eng}.
The app's privacy policy and specifications are available on the Japanese government website\cite{ref:shiyou}.

The Japanese government announced that there have been cases where COVID-19 symptoms do not appear\cite{musho}.
Roth et al.\cite{roth} reported that even if the infectors of COVID-19 are in their incubation period, it is possible that they can infect others.
Consequently, it is possible that infectors, who do not show symptoms, can infect many persons.
In general, users can know the personal information of infectors via the app.
If infectors continue to stay at home, he/she could avoid infecting others.
Therefore, the following 3 points are important to overcome the spreading of COVID-19 by utilizing the app:
(1) use the App; 
(2) users who know about their contact with infectors via the app stay at home, 
(3) infectors register that they are infected via the app. 
Moreover, other countries (e.g., China, India, Israel, and so on) also have developed similar apps\cite{ref:cont}.
However, to overcome the spreading of COVID-19, the target values of the described 3 points are unclear.

In the real world, we cannot strictly verify the effectiveness of the app because there are various mixed measures for overcoming the spread of COVID-19. 
Moreover, since human life and death are involved, we cannot conduct a control experiment in the real world owing to infection spread and convergence.
Thus, we had better conduct experiments in an artificial world, and one of the approaches is applying multi-agent simulation (MAS).
There are numerous previously conducted studies based on virus-spreading simulations \cite{ref_mas1, ref_mas2, ref_mas3, SEIR_MAS}.
However, these studies do not target COVID-19, and we cannot verify the app's effectiveness.

Very recently, simulations targeting how to overcome the spread of COVID-19 spreading have been conducted in the year 2020.
For example, by applying the susceptible-exposed-infectious-recovered (SEIR) model to study COVID-19, Hou et al.\cite{SEIR_C19_1} observed that a measure of decreasing contact with persons can effectively decrease the total number of infectors at peak time.
Chatterjee et al.\cite{SEIR_C19_3} also developed a SEIR model for COVID-19 and conducted a simulation experiment using India as a case study.
As a result, these authors reported that the measures of avoiding contact with persons, such as lockdown, can significantly reduce the spread of COVID-19. 

These studies are beneficial.
Besides, the described SEIR model-based COVID-19 simulations \cite{SEIR_C19_1, SEIR_C19_3}  do not involve the app's effect.
To survey the app's effectiveness using England as a case study, Hinch et al.\cite{tr1} conducted simulations based on a mixed method of SIR model (it is not a SEIR model) and agent-approach simulation.
They concluded that the usage of the app by 56\% of the total population can lead to the convergence of COVID-19.
Although this is an innovative result, it is difficult to verify its reliability in the case of social simulations.

In general, the reliability of simulations is measured by calculating the difference between the data collected in real world and those generated from a simulation model.
In addition, Takahashi\cite{ref:rel} pointed out that this method is not always possible in the case of social simulation.
In the case of the COVID-19 pandemic situation, it is difficult to collect data to calculate reliability in the real world.
Therefore, we cannot measure the reliability of COVID-19 spreading and the app's effectiveness by calculating the difference between data of the real world and simulator. 
As an alternative method, Takahashi\cite{ref:rel} recommended the comparison of results generated from different simulation models.
If different simulation models generated similar results, then we can justify that the simulation result is reliable.
Further, COVID-19 and the app, such as COCOA, are the latest research topics of the year 2020 (COCOA developed by the Japanese government was released on Jun. 19, 2020).
Therefore, simulators for verifying the effectiveness of the apps, such as COCOA, are very few as of Aug. 2020.
Thus, applying a measuring method of reliability based on the comparison of different simulation models is difficult. 
In conclusion, we observe that now is the time to report numerous simulation cases based on the effectiveness of the app for the convergence of COVID-19 for different methods.
In the future, the accumulation of different case studies by various researchers can enable the study of the measurement of simulation reliability.

Therefore, by improving a simulator developed by Omae et al.\cite{ref:omae}, we introduce a multi-agent simulator that can express COVID-19 spreading and the usage of the app, such as COCOA.
To overcome the spreading of COVID-19, we employ the following three points:
(1) install and use the app; 
(2) users who have contact with infectors via the app stay at home; 
(3) infectors register that they have been infected via the app. 

Furthermore, we include three parameters for the following expression:
(1') the usage rate of the app; 
(2') decreasing value of going out probability of persons who have contact with infectors via the app; 
(3') infection registration rate of infectors through the app.
In this study, we introduce the details of the developed MAS and the app's effectiveness of reducing the number of infectors of COVID-19.

\vspace{-10pt}
\section{Simulator}\label{sec:pre}
In this section, we describe the multi-agent simulator for the spreading of the virus infectors developed by Omae et al.\cite{ref:omae}.
Moreover, the new parameters for expressing the app embedded in the simulator are described.

\begin{figure}[t]
\centering
\includegraphics[scale=0.7]{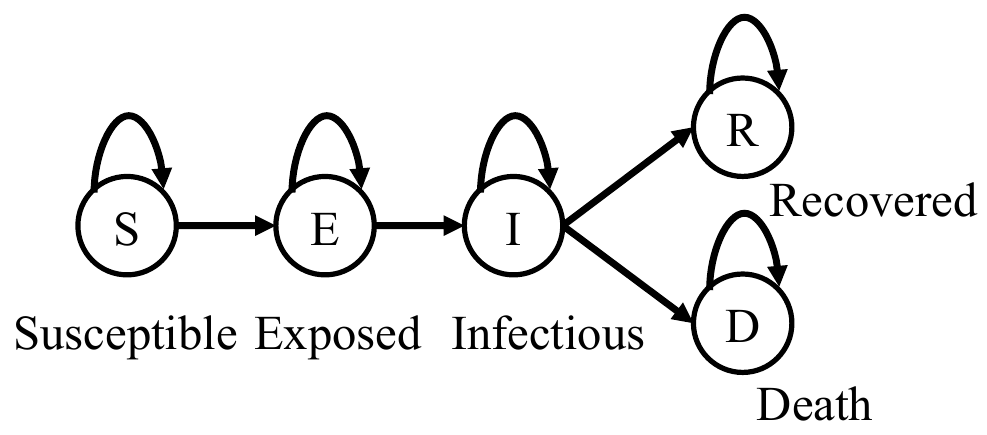}
\caption{Infection transition model.}
\label{fig:automa}
\end{figure}

\vspace{-10pt}
\subsection{Transition of infection: The SEIR model}\label{sec:SEIR}
The SEIR model is one of the methods for simulating the spread of the virus.
There are four states (S, E, I, R) in the model \cite{SEIR_C19_1}.
Furthermore, Susceptible person (State: S) has a possibility of being an infector by having contact with other infectors.
Exposed person (State: E) means an incubation period.
Infectious person (State: I) means an infector of a virus.
Recovered person (State: R) means the person who has recovered and acquired immunity.
In the case of SEIR model, there have been some simulations of COVID-19 \cite{SEIR_C19_1, SEIR_C19_2, SEIR_C19_3, SEIR_C19_4, SEIR_C19_5}. 
Thus, we apply the SEIR model for the simulations of COVID-19 spreading.
We note that the basic SEIR model is in the form of an ordinary differential equation.
It can be fast simulated, but including various parameters, such as the expression of COCOA, agents' job, lifestyle, and so on, is difficult.
Moreover, a mixed model of the SEIR model and MAS has been proposed \cite{SEIR_MAS, ref_mas1}.
This is the model of a virus that stochastically spread through the contact of agents.
In this study, we use a mixed model of the SEIR model and MAS.

An infection transition model is depicted in {\bf Fig.\ref{fig:automa}}.
The arrows denote a possibility of states' transition and give transition probabilities.
If no arrow exists, states' transition cannot occur.
Additionally, states' transitions are from S to E, from E to I, and from I to R or D.
In the basic SEIR model, state D does not exist.
State D means death.
The MAS developed by Omae et al.\cite{ref:omae} included state D because they conducted a simulation to verify the relationship between the number of dead persons and the capacity limit of isolation wards.

Next, we explain transition probabilities.
The stochastic variable for the transition of S, E, I, R, D is defined by
\begin{equation}
P(X_{t+1} | X_{t}, C, T, H). \label{e1}
\end{equation}
If $C, T, H$ are omitted, then the transition probability has the same value whatever value we assigned to the variable.
Here, $X_t$ and $X_{t+1}$ denote the infection state of an agent at time $t$ and $t+1$, and they are defined by
\begin{equation}
X_t, X_{t+1} \in \{ {\mathrm S}, {\mathrm E}, {\mathrm I}, {\mathrm R}, {\mathrm D}  \}.
\end{equation}
Moreover, $C$ denotes the variable for expressing contact with agents of state I (infectors);
$C=0$ denotes not-contact; $C=1$ means contact with agents of state I (infectors).
$T$ denotes the number of days elapsed from changing to other states (a unit is [days]).
$H$ denotes the variable for expressing hospitalization of agents of state I ($H=0$: nonhospitalization; $H=1$: hospitalization).
In addition, Eq.(\ref{e1})-based transition implies that transition probabilities depend on $C$, $T$, and $H$.
The strict meaning of contact with an agent and infectors is that they are in contact within 1 Euclidean distance for about 10 minutes in an artificial society.

First, the transition probability from state S to state S is defined by
\begin{equation}
P(X_{t+1}={\mathrm S} | X_{t}={\mathrm S}, C) =
  \begin{cases}
    1, & (C=0), \\
    1-\beta, & (C=1),
  \end{cases}
\end{equation}
where $\beta$ denotes an infection probability of about 10 minutes (the minimum unit of time of the simulator is 10 minutes\cite{ref:omae}).
In the case in which there is no contact with a target agent and infectors, the transition probability is zero.
Further, in the case of contact with them, the transition probability is decreased.

The transition probability from state S to state E is defined as follows:
\begin{equation}
P(X_{t+1}={\mathrm E} | X_{t}={\mathrm S}, C) =
  \begin{cases}
    0, & (C=0), \\
    \beta, & (C=1).
  \end{cases} \label{eq:se}
\end{equation}
If a target agent had contact with infectors ($C=1$), the transition probability from state S to state E is $\beta$.
Since state S only changes to state S or E, we have that
\begin{equation}
P(X_{t+1}={\mathrm S} | X_{t}={\mathrm S}) + P(X_{t+1}={\mathrm E} | X_{t}={\mathrm S}) = 1
\end{equation}
is satisfied for all conditions of $C, T, H$.

The transition probability from state E to state E is defined as follows:
\begin{equation}
P(X_{t+1}={\mathrm E} | X_{t}={\mathrm E}, T) =
  \begin{cases}
    0, & (T = T_{\mathrm E \rightarrow \mathrm I}), \\
    1, & (T \neq T_{\mathrm E \rightarrow \mathrm I}),
  \end{cases}
\end{equation}
while the transition probability from state E to state I is defined as follows:
\begin{equation}
P(X_{t+1}={\mathrm I} | X_{t}={\mathrm E}, T) =
  \begin{cases}
    0, & (T \neq T_{\mathrm E \rightarrow \mathrm I}), \\
    1, & (T = T_{\mathrm E \rightarrow \mathrm I}),
  \end{cases}
\end{equation}
where $T_{\mathrm E \rightarrow \mathrm I}$ denotes the period of required transition time [day] (incubation period of virus). 
If $T$ reaches $T_{\mathrm E \rightarrow \mathrm I}$, the state of an agent changes from state E to state I.
In the presence of other conditions, the target agent remains in state E.
Since state E only changes to state E or I, we obtain that
\begin{equation}
P(X_{t+1}={\mathrm E} | X_{t}={\mathrm E}) + P(X_{t+1}={\mathrm I} | X_{t}={\mathrm E}) = 1
\end{equation}
is satisfied for all conditions of $C, T, H$.

The transition probability from state I to state I is defined by
\begin{equation}
P(X_{t+1}={\mathrm I} | X_{t}={\mathrm I}, T) =
  \begin{cases}
    1, & (T \neq T_{\mathrm I \rightarrow \mathrm{RD}}), \\
    0, & (T = T_{\mathrm I \rightarrow \mathrm{RD}});
  \end{cases}
\end{equation}
the transition probability from state I to state R is defined by
\small
\begin{equation}
P(X_{t+1}={\mathrm R} | X_{t}={\mathrm I}, T, H) =
  \begin{cases}
    0, & (T \neq T_{\mathrm I \rightarrow \mathrm{RD}} \land H=0), \\
    1-\gamma_0, & (T = T_{\mathrm I \rightarrow \mathrm{RD}} \land H = 0), \\
    0, & (T \neq T_{\mathrm I \rightarrow \mathrm{RD}} \land H=1), \\
    1-\gamma_1, & (T = T_{\mathrm I \rightarrow \mathrm{RD}} \land H = 1). 
  \end{cases}\label{imp_e1}
\end{equation}
\normalsize
In addition, the transition probability from state I to state D is defined by
\small
\begin{equation}
P(X_{t+1}={\mathrm D} | X_{t}={\mathrm I}, T, H) =
  \begin{cases}
    0, & (T \neq T_{\mathrm I \rightarrow \mathrm{RD}} \land H=0), \\
    \gamma_0 & (T = T_{\mathrm I \rightarrow \mathrm{RD}} \land H = 0), \\
    0, & (T \neq T_{\mathrm I \rightarrow \mathrm{RD}} \land H=1), \\
    \gamma_1, & (T = T_{\mathrm I \rightarrow \mathrm{RD}} \land H = 1), 
  \end{cases}\label{imp_e2}
\end{equation}
\normalsize
where $T_{\mathrm I \rightarrow \mathrm{RD}}$ denotes the period of required transition time [day] (infection period of virus). 
If $T$, which is the number of days elapsed from changing to state I, reaches $T_{\mathrm I \rightarrow \mathrm{RD}}$, then the state of an agent changes from state I to state R or D.
In the case in which other conditions are present, the target agent maintains state I.
Additionally, $\gamma_0$ represents the fatality rate in the case of nonhospitalization ($H=0$), while $\gamma_1$ represents the fatality rate in the case of hospitalization ($H=1$).
Further, the fatality rate depends on the hospitalization of agents.
In general, because the fatality rate in the case of nonhospitalization has higher value more than that of hospitalization, we recommend
\begin{equation}
\gamma_0 > \gamma_1
\end{equation}
as the fatality rates.
Since state I only changes to state I, R, or D, the following equation
\begin{equation}
\sum_{x \in \{{\mathrm I}, {\mathrm R}, {\mathrm D}\}} P(X_{t+1}=x | X_{t}={\mathrm I}) = 1
\end{equation}
is satisfied for all conditions of $C, T, H$.

The state R denotes that the agents acquired immunity, and the state D means death.
Therefore, the following equations are satisfied:
\begin{equation}
P(X_{t+1}={\mathrm R} | X_{t}={\mathrm R}) =  1,
\end{equation}
\begin{equation}
P(X_{t+1}={\mathrm D} | X_{t}={\mathrm D}) =  1.
\end{equation}

The described transition probabilities were proposed by Omae et al.\cite{ref:omae}.
In this study, we also use them for infection simulations.

\tabcolsep = 1pt
\begin{table}[tb]
\caption{The simulator's parameters of a previous study\cite{ref:omae}}\label{tab1}
\vspace{-10pt} 
\begin{center}
\begin{tabular}{cll}\hline
\multicolumn{1}{c}{Variables} & \multicolumn{1}{c}{parameters} & \multicolumn{1}{c}{values} \\ \hline
$e_{1}$&Max simulation period& days\\ %
$e_{2}$&The number of houses& numbers\\ %
$e_{3}$&The number of initial infectors& persons \\ %
$e_{4}$&Locations of companies&XY\\ %
$e_{5}$&Locations of shops&XY\\ %
$e_{6}$&Locations of schools&XY\\ %
$e_{7}$&Capacity limitation of isolation wards& beds\\ %
$e_{8}$&Basic going out probability& prob. \\ %
$e_{9}$&Going out time& time \\ %
$e_{10}$&Stay time of facility& time \\ %
$e_{11}$&Probability of going to a hospital& prob.\\ %
$e_{12}$&Decreasing value of $e_8$ during state I& prob. \\ \hline
$\beta$&Infection probability& prob.\\ %
$\gamma_0$&Fatality rate (nonhospitalization) & prob. \\ %
$\gamma_1$&Fatality rate (hospitalization) & prob.\\
$T_{\mathrm E \rightarrow \mathrm I}$ & Incubation periods (from E to I) &days \\
$T_{\mathrm I \rightarrow \mathrm{RD}}$ & Infection periods (from I to R or D) &days\\ \hline
\end{tabular}
\end{center}
\end{table}

\vspace{-10pt} 
\subsection{Simulations flow}\label{pre2}
First, we describe parameters of agents, simulation environment, and their initial conditions.
The agents live in a 2-dimensional (2D) space ($x$ and $y$ axes) with minimum and maximum values 0 and 1000 as an artificial society.
The agents' locations are expressed by utilizing 2D real numbers from 0 to 1000.
The parameters for expressing them are shown in {\bf Table \ref{tab1}}.
Additionally, the max simulation period is denoted by $e_1$, 
and the number of houses is denoted by $e_2$.
We assume that 3 persons (an office worker, homemaker, and student) live in a house.
Therefore, the total number of populations in an artificial society is $3 \times e_2$.
Next, the number of initial infectors (agents of initial state I) is denoted by $e_3$.
The state of $e_3$ persons out of $3e_2$ persons becomes state I (infection).
Besides, the state of other persons becomes state S.
Afterward, $(x, y)$ coordinates of the agents' houses and destination facility locations (company, shop, or school) are decided.
A coordinate of a house means the location that agents live daily, and its number is $e_2$.
The destination facility location means the place an individual goes almost daily. It is possible that agents may not go there.
The facilities for office workers, homemakers, and students are company, shop, and school, respectively.
In a society, there are numerous companies, shops, and schools.
Therefore, we choose many $(x, y)$ coordinates as variables of $e_{\{4,5,6\}}$.
In the case of office workers, the company to choose is decided by a uniform random number at simulation start timing.
As with office workers, destination facility locations of homemakers and students are selected.
Further, the destination facility location is one per agent.
After the decision, they do not change.

Moreover, it is possible that agents of state I do not go to the destination facility, such as a company or school, while they go to a hospital to be hospitalized.
However, if the capacity limit of isolation wards is reached, they may not be hospitalized even if the agent is an infector. 
Consequently, the capacity limit of isolation wards denoted by $e_7$ is decided.
Moreover, we set parameters of $\beta, \gamma_0, \gamma_1, T_{\mathrm E \rightarrow \mathrm I}, T_{\mathrm I \rightarrow \mathrm{RD}}$ described in Subsection \ref{sec:SEIR}.

After setting the initial parameters, we start our simulation.
Further, there are ``1-day process'' and ``1-step process''.
We remark that 1-day process means that the process is conducted at the start timing of a day.
In this process, whether or not agents go to the destination facility location (company, shop, and school) is decided based on the basic going out probability $e_8$.
This value is different for each of the agents.
If the agents go to the destination facility location, the going out time $e_9$ and stay time of facility $e_{10}$ are decided.
However, the agents of state I go to hospital depending on the ``probability of going to a hospital'' $e_{11}$.
If the capacity limit of isolation wards is not exceeded, the agents of state I are hospitalized.
In this case, the basic going out probability $e_8$ of the agents becomes zero.
However, if the capacity limit of isolation wards is exceeded, the agents of state I are not hospitalized.
In this case, agents of state I are not isolated in a hospital.
Therefore, it is possible that they can go outside even if they are infected.
Then, since they may feel sick, the basic going out probability $e_8$ is reduced by ``decreasing value during state I'' $e_{12}$.

Besides, the 1-step process means that the process is repeatedly conducted at each minimum unit of time of a simulation.
In our simulator, the minimum unit of time is 10 minutes.
Therefore, 1 step is equal to 10 minutes.
Agents who decide to go to the facility by a 1-day process can go out, whereas other agents stay at home.
Thus, office workers, homemakers, and students go to company, shop, and school, respectively.
Locations of companies, shops, and schools are respectively denoted by $e_{\{4, 5, 6\}}$, and their coordinates are destinations of agents.
The going out time to destination is denoted by $e_9$; at this time, agents go there in the available shortest Euclidian distance.
After arriving at the destination, agents stay there based on stay time of facility $e_{10}$.
Afterward, agents go back to their homes.

The described process is called a ``1-step process.''
Since 1 step is 10 minutes, 1 day is equal to 144 steps (24 hours).
After completing 144 steps, another day is repeatedly started.
When the max simulation period $e_1$ is reached, a simulation is completed.

\begin{figure}[t]
\centering
\includegraphics[scale=0.7]{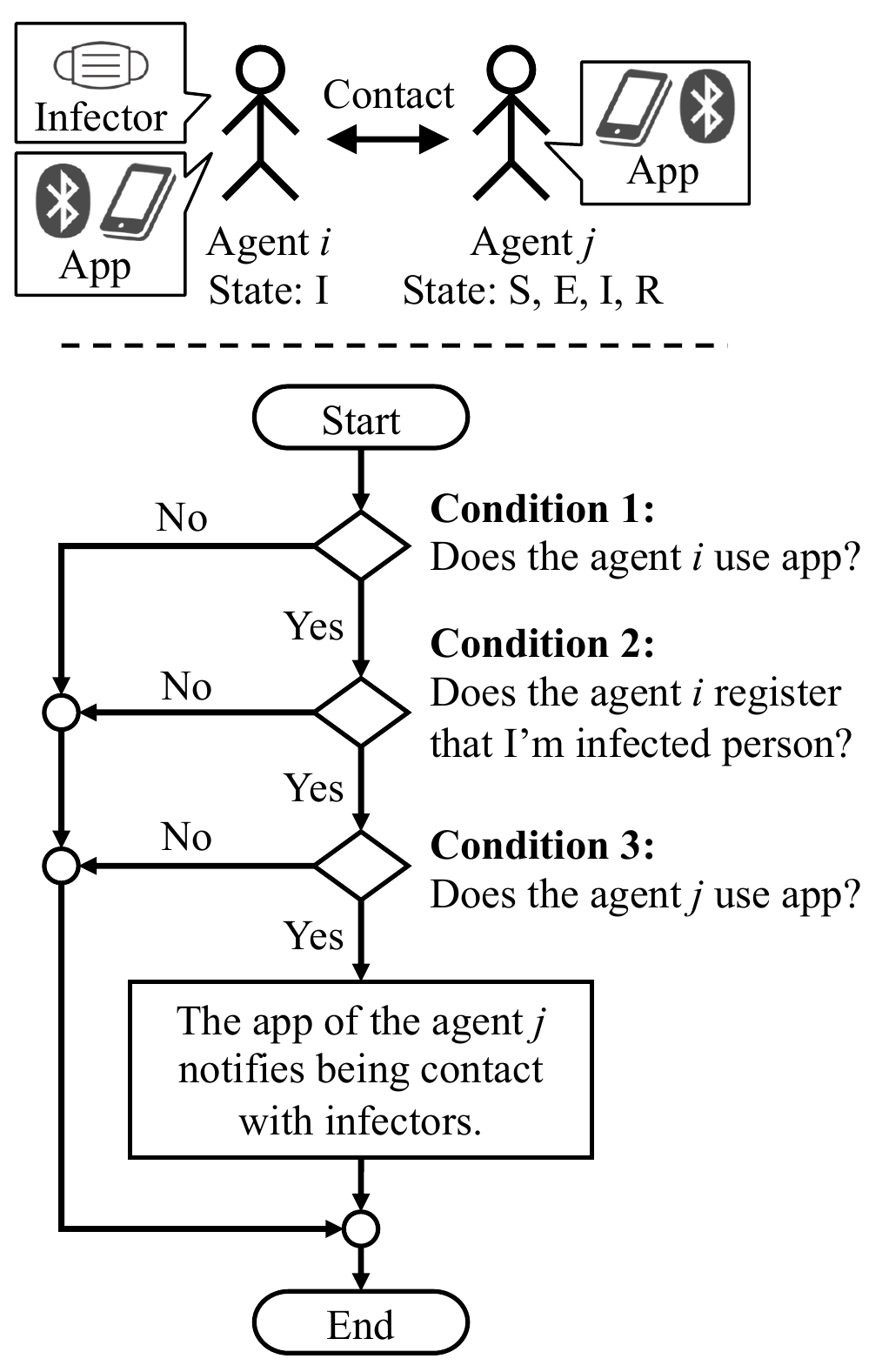}
\caption{A notification process of the app in the case of contact with infectors and other agents.}
\label{fig:notice}
\end{figure}

\vspace{-10pt} 
\subsection{Contact-confirming application in MAS}
This study aims to verify the effectiveness of COCOA \cite{ref:COCOA_eng} by applying MAS.
Therefore, we establish parameters to incorporate the app into the existing simulator developed by Omae et al.\cite{ref:omae} described in Subsections \ref{sec:SEIR} and \ref{pre2}.
The newly included parameter is
\begin{equation}
\bm{p}_\mathrm{app} = (p_\mathrm{app}^{1}, p_\mathrm{app}^{2}, p_\mathrm{app}^{3}), \label{eq:cocoa}
\end{equation}
and we call it ``app parameter'' because it is used to express COCOA.
Here, 
$p_\mathrm{app}^1$ denotes the usage rate of the app for all the population;
$p_\mathrm{app}^2$ denotes the decreasing value of going out probability of agents that received notification of their contact with infectors from the app (infectors mean agents of state I);
$p_\mathrm{app}^3$ denotes the infection registration rate of infectors to the app.

A notification process of the app, such as COCOA, to an agent that came in contact with infectors is illustrated in {\bf Fig.\ref{fig:notice}}.
As depicted in {\bf Fig.\ref{fig:notice}}, agent $i$ is of state I (infector) and agent $j$ is the person that received notification of having contact with infectors from the app.
Condition 1 checks whether or not agent $i$ uses the app.
If condition 1 is satisfied, condition 2 checks whether or not agent $i$ registered that he/she is an infector.
Conditions 1 and 2 are conditional branches of the infectors' side.
If condition 2 is true, then condition 3 checks whether or not the agent $j$ utilizes the app.
Furthermore, if conditions 1, 2, and 3 are all satisfied, the app of agent $j$ notifies about having contact with an infector or infectors.
Consequently, agent $j$ can know that ``he/she had contact with an infected person.''
Therefore, the basic going out probability of agent $j$ is reduced by app parameter $p_\mathrm{app}^2$.
The higher the $p_\mathrm{app}^1$, the more likely conditions 1 and 3 are satisfied.
Moreover, the higher the $p_\mathrm{app}^3$, the more likely condition 2 is satisfied.
Persons who come in contact with infectors should not be allowed to go out.
Besides, if they do not properly use the app, they may not be aware of having contact with an infector or infectors.
Thus, if each element of the app parameters $\bm{p}_\mathrm{app}$ is increased, then we believe that the number of infectors could be decreased.

In an artificial society, the reducing period of going out probability of agent $j$ who got contact information via the app is 2 weeks.
This is because the contact recording period of the real app developed by the Japanese government in the real world is 2 weeks \cite{ref:shiyou}. 
Moreover, the real app notifies about contact information when a person has contact with infectors for at least 15 minutes \cite{ref:shiyou}.
Further, our simulator's minimum unit time is 10 minutes.
Therefore, if agents have contact with infectors for at least 10 minutes, then the app notifies them concerning this.

\tabcolsep = 3pt
\begin{table*}[tb]
\caption{Simulation conditions}\label{tab2}
\vspace{-10pt} 
\begin{center}
\begin{tabular}{cll}\hline
\multicolumn{1}{c}{Variables} & \multicolumn{1}{c}{parameters} & \multicolumn{1}{c}{values} \\ \hline
$e_{1}$&Max simulation period& 45 [day]\\ %
$e_{2}$&The number of houses& 333 [houses] (999 [ppl])\\ %
$e_{3}$&The number of initial infectors& 10 [ppl] \\ %
$e_{4}$&Facility locations (companies) & 3 places: $(x, y)= (200, 800), (500, 500) , (800, 100)$\\ %
$e_{5}$&Facility locations (shops)& 3 places: $(x, y)= (200, 500), (500, 100) , (800, 800)$\\ %
$e_{6}$&Facility locations (schools)& 3 places: $(x, y)= (200, 100), (500, 800) , (800, 500)$\\ %
$e_{7}$&Capacity limitation of isolation wards& 0 [beds]\\ %
$e_{8}$&Basic going out probability (office worker)&99.0 $\sim$ 100.0 [\%] \\ %
$e_{8}$&Basic going out probability (homemaker) & 50.0 $\sim$ 100.0 [\%]\\ %
$e_{8}$&Basic going out probability (student) & 99.0 $\sim$ 100.0 [\%]\\ %
$e_{9}$&Going out time (office worker) & $8:30 \pm 1:30$\\ %
$e_{9}$&Going out time (homemaker) & $10:30 \pm 1:30$ \\ %
$e_{9}$&Going out time (student) & $8:30 \pm 1:30$ \\ %
$e_{10}$&Stay time of facility (office worker) & 6:00 $\sim$ 8:00\\ %
$e_{10}$&Stay time of facility (homemaker)& 0:10 $\sim$ 0:30 \\ %
$e_{10}$&Stay time of facility (student) & 5:00 $\sim$ 6:00 \\ %
$e_{11}$&Probability of going to a hospital& - [\%]\\ %
$e_{12}$&Decreasing value of $e_8$ during state I& 30.0 [\%]\\ \hline
$\beta$&Infection probability& 0.006 [\%]\\ %
$\gamma_0$&Fatality rate (nonhospitalization) & 10.0 [\%]\\ %
$\gamma_1$&Fatality rate (hospitalization) & - [\%]\\
$T_{\mathrm E \rightarrow \mathrm I}$ & Incubation periods (from E to I) &3, 5, 7 [day] \\
$T_{\mathrm I \rightarrow \mathrm{RD}}$ & Infection periods (from I to R, D) &8, 10, 12 [day]\\ \hline
$p_\mathrm{app}^1$& Usage rate of the app& 0, 20, $\cdots$, 100 [\%]\\
$p_\mathrm{app}^2$& DVP* during contact notification from the app &0, 20, $\cdots$, 100  [\%] \\
$p_\mathrm{app}^3$& Registration rate of infected persons &0, 20, $\cdots$, 100  [\%]\\ \hline
\multicolumn{3}{r}{$a \sim b$: uniform random number from $a$ to $b$.}\\
\multicolumn{3}{r}{$a \pm b$: gaussian random number of mean $a$ and std. $b$.}\\
\multicolumn{3}{r}{DVP*: Decreasing value of going out probability}\\
\end{tabular}
\end{center}
\vspace{-10pt} 
\end{table*}

\vspace{-10pt} 
\section{Experiment}
\subsection{Experimental objectives and conditions}
We examine that the app can effectively reduce the number of infectors.
Therefore, we conduct simulations to verify the effectiveness of the app, such as COCOA.
Besides, our simulations' conditions are presented in {\bf Table \ref{tab2}}.
The max simulation period $e_1$ is 45 days, and the number of houses $e_2$ is 333 (the total population is 999).
The number of initial infectors $e_3$ is 10 persons.
Further, the basic going out probability $e_8$, going out time $e_9$, and stay time of facility $e_{10}$ of each agent are decided by uniform random number or Gaussian random number. 
A method for giving parameters of going out by applying probability distributions is based on previously conducted studies of mixed models of MAS and SEIR\cite{SEIR_MAS, ref:omae}.
The number of companies, shops, and schools is 3 facilities (the number of total facilities is 9).

Next, we describe the parameters of the infection transition.
We recall that the minimum unit of time of the simulator is 10 minutes. 
Therefore, it is desirable to set an actual COVID-19 infection probability $\beta$ for 10 minutes.
However, an actual COVID-19 infection probability is unclear.
Consequently, we find the infection probability that persons from 5\% to 10\% of the total population become infectors in 45 days when all agents do not use the app.
For this result, we set $\beta = 0.006\%$ as the infection probability.

Afterward, we set parameters $T_{\mathrm E \rightarrow \mathrm I}$ and $T_{\mathrm I \rightarrow \mathrm{RD}}$.
Ohashi's assumption\cite{ref:ohashi} states that the average incubation period is 5 days and the average infection period is 10 days.
Parameters $T_{\mathrm E \rightarrow \mathrm I}$ and $T_{\mathrm I \rightarrow \mathrm{RD}}$ were determined by adding $\pm 2$ days to Ohashi's assumption\cite{ref:ohashi}.
Thus, the incubation periods $T_{\mathrm E \rightarrow \mathrm I}$ are 3, 5, and 7 days.
The infection periods $T_{\mathrm I \rightarrow \mathrm{RD}}$ are 8, 10, and 12 days.
Parameters $T_{\mathrm E \rightarrow \mathrm I}$ and $T_{\mathrm I \rightarrow \mathrm{RD}}$ are decided by uniform random number from the above dates for each agent.
Since we consider that agents in state I are sick, their going out probability decreases.
We set 30\% as the decreasing value of going out probability of state I's agents $e_{12}$.
To confirm the effectiveness of reducing the number of infectors of only the App, there are no hospitals in an artificial society.
Therefore, we set 0 as the capacity limit of isolation wards $e_7$, and this implies that agents cannot be admitted to hospitals.
As a result, the probability of going to a hospital $e_{11}$ and fatality rate (hospitalization) $\gamma_1$ in {\bf Table \ref{tab2}} are empty.
Next, we consider fatality rate (nonhospitalization) $\gamma_0$.
For COVID-19's fatality rate, the Mitsubishi Research Institute \cite{mitsu} reported that the fatality rate of countries that were medically collapse is over 10\%.
Even though in other cases, the fatality rate is about 1\% (e.g., Italy: 14.3\%, Spain: 11.3\%, Iceland: 0.6\%, Singapore: 1.1\%, as of Apr. 2020 \cite{mitsu}. In the case of Italy, the death of out-of-hospital increased because of the COVID-19 outbreak according to Baldi et al.\cite{ref:baldi}).
Therefore, we set 10\% as fatality rate (nonhospitalization) $\gamma_0$.

Next, we describe the app's parameters $\bm{p}_\mathrm{app}$ defined by Eq.(\ref{eq:cocoa}).
Our study objective is to verify the effectiveness of the app. 
Therefore, we set many combinations as the app parameters.
As demonstrated in {\bf Table \ref{tab2}}, the values of $p_\mathrm{app}^1$, $p_\mathrm{app}^2$ and  $p_\mathrm{app}^3$ are $0\%, 20\%, \cdots, 100\%$, respectively.
Since the value of a parameter is 6 patterns, all parameters' combination is $6\times6\times6 = 216$ patterns.
Additionally, because the simulations of infection spread are stochastic events, it is desirable to conduct simulation multiple times per scenario and calculate the average value of the total infectors.
If this is not the case, the consequences can be influenced by chance.
Therefore, we conduct simulations of about 30 times per scenario while changing the random seeds.
In other words, the total number of conducted simulation for 45 days is $216$ scenarios $\times$ 30 random seeds $= 6480$ times.

\vspace{-10pt} 
\subsection{Results and discussions: at the end of simulations} \label{sub32}
Now, we checked the number of total infectors of the scenario of all agents who do not use the app.
In the results generated by random seed of 30 patterns, there were 2 cases in which the infection did not spread even though the app was not used (the total number of infectors at the end of the simulation is below 30).
Since the random seeds generating the above results are inappropriate to verify the effectiveness of the app, we excluded them from our analysis.

\begin{figure}[t]
\centering
\includegraphics[scale=0.6]{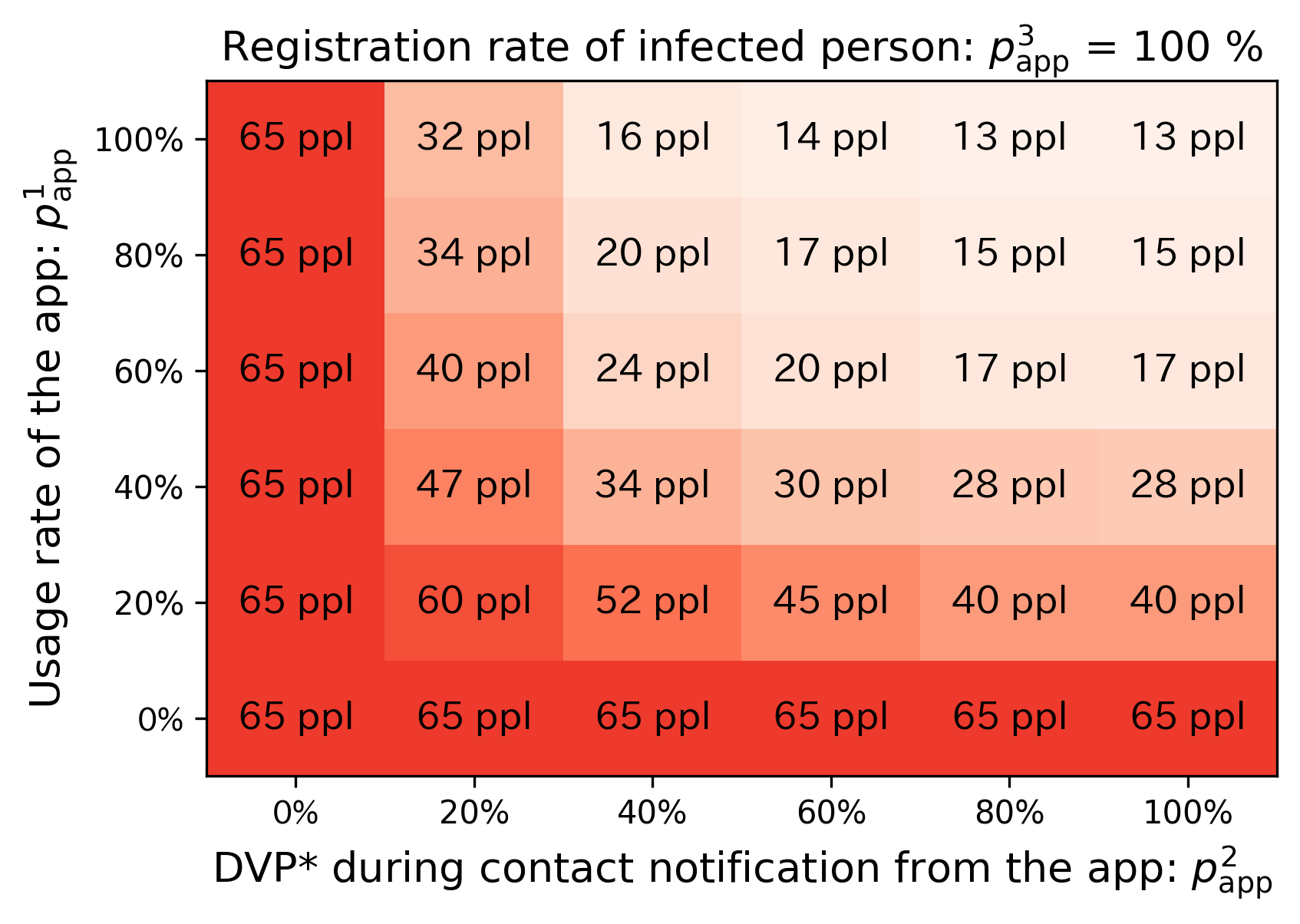}
\caption{The relationship between the app parameters $p_\mathrm{app}^{\{1, 2, 3\}}$ and the number of total infectors [ppl] at the end of the 45 days simulations (in the case of $p_\mathrm{app}^{3}=100\%$). The higher the number of infectors, the redder. DVP*: Decreasing value of going out probability.}
\label{fig:1map}
\end{figure}

Thereafter, we calculated the total number of infectors.
The total number of infectors means the sum value of state E, I, R, and D.
Further, the total number of infectors at the end of the 45 days simulation of the scenarios of all infectors who registered that they are infected (i.e., $p_\mathrm{app}^{3}=100\%$) is shown in {\bf Fig.\ref{fig:1map}}.
The vertical and horizontal axes denotes the respective app parameters $p_\mathrm{app}^{\{1, 2\}}$.
If either $p_\mathrm{app}^{1}$ or $p_\mathrm{app}^{2}$ is 0, it implies that the app does not work.
Therefore, the number of infectors in row and column of $p_\mathrm{app}^{\{1, 2\}}=0\%$ has the same value (65 persons).
Additionally, this number is less than 65 if the app is effective.
As illustrated in {\bf Fig.\ref{fig:1map}}, as $p_\mathrm{app}^{1}$ and $p_\mathrm{app}^{2}$ increase, the total number of infectors decreases. 

From the result presented in {\bf Fig.\ref{fig:1map}}, we consider the target value of the app usage strategy required to halve the number of infectors compared with when the app is not used.
Since the total number of infectors is 65 persons in the case in which the app is not used, the standard value is less than $65\times(1/2)=32.5$ persons.
As shown in {\bf Fig.\ref{fig:1map}}, there are no cases of the number of infectors less than 32.5 persons in the result of the usage rate of the app ($p_\mathrm{app}^{1}=20\%$). 
Therefore, the usage rate of app $p_\mathrm{app}^{1}$ has to be at least 40\%.
In the case in which $40\% \geq p_\mathrm{app}^{1}$, the scenarios that the total number of infectors is less than 32.5 persons are as follows:
\begin{eqnarray}
(p_\mathrm{app}^{1}, p_\mathrm{app}^{2}, p_\mathrm{app}^{3}) &=& (40\%, 60\%, 100\%), \label{s1}\\
(p_\mathrm{app}^{1}, p_\mathrm{app}^{2}, p_\mathrm{app}^{3}) &=& (60\%, 40\%, 100\%),  \label{s2}\\
(p_\mathrm{app}^{1}, p_\mathrm{app}^{2}, p_\mathrm{app}^{3}) &=& (100\%, 20\%, 100\%)\label{s3}.  
\end{eqnarray}
For all populations, it is difficult to reach the usage rate ($100\%$) of the app $p_\mathrm{app}^{1}$.
Therefore, $\bm{p}_\mathrm{app}=(40\%, 60\%, 100\%)$ or $(60\%, 40\%, 100\%)$ is a realistic target value.
We confirmed that the total number of infectors become under half value If about half of agents use the App and about half of the frequency of going to the facility.

Next, we consider the target value for reducing the number of infectors to about 2/3.
The standard value is less than $65\times(2/3)=43.3$ persons.
From the result depicted in {\bf Fig.\ref{fig:1map}}, the scenarios that the total number of infectors is less than 43.3 persons are as follows:
\begin{eqnarray}
(p_\mathrm{app}^{1}, p_\mathrm{app}^{2}, p_\mathrm{app}^{3}) &=& (20\%, 80\%, 100\%),  \\
(p_\mathrm{app}^{1}, p_\mathrm{app}^{2}, p_\mathrm{app}^{3}) &=& (40\%, 40\%, 100\%), \\
(p_\mathrm{app}^{1}, p_\mathrm{app}^{2}, p_\mathrm{app}^{3}) &=& (60\%, 20\%, 100\%). 
\end{eqnarray}
The first case $\bm{p}_\mathrm{app} = (20\%, 80\%, 100\%)$ means that if the usage rate of the app is low, the agents who had being in contact with infectors extremely reduce their going out.
The second and third cases $\bm{p}_\mathrm{app} = (40\%, 40\%, 100\%)$ and $(60\%, 20\%, 100\%)$ are the relaxation of Eq.(\ref{s1})--(\ref{s3}), which can reduce the number of infectors by half.
As of Aug. 2020, the Japanese population download rate of COCOA is about 10\%\cite{ref:6p}.
Therefore, the usage rate of app $p_\mathrm{app}^{1}=20\%$ is criteria that will be reached in the near future.
Consequently, the most important scenario is the first case $\bm{p}_\mathrm{app} = (20\%, 80\%, 100\%)$.

\begin{figure*}[t]
\centering
\includegraphics[scale=0.7]{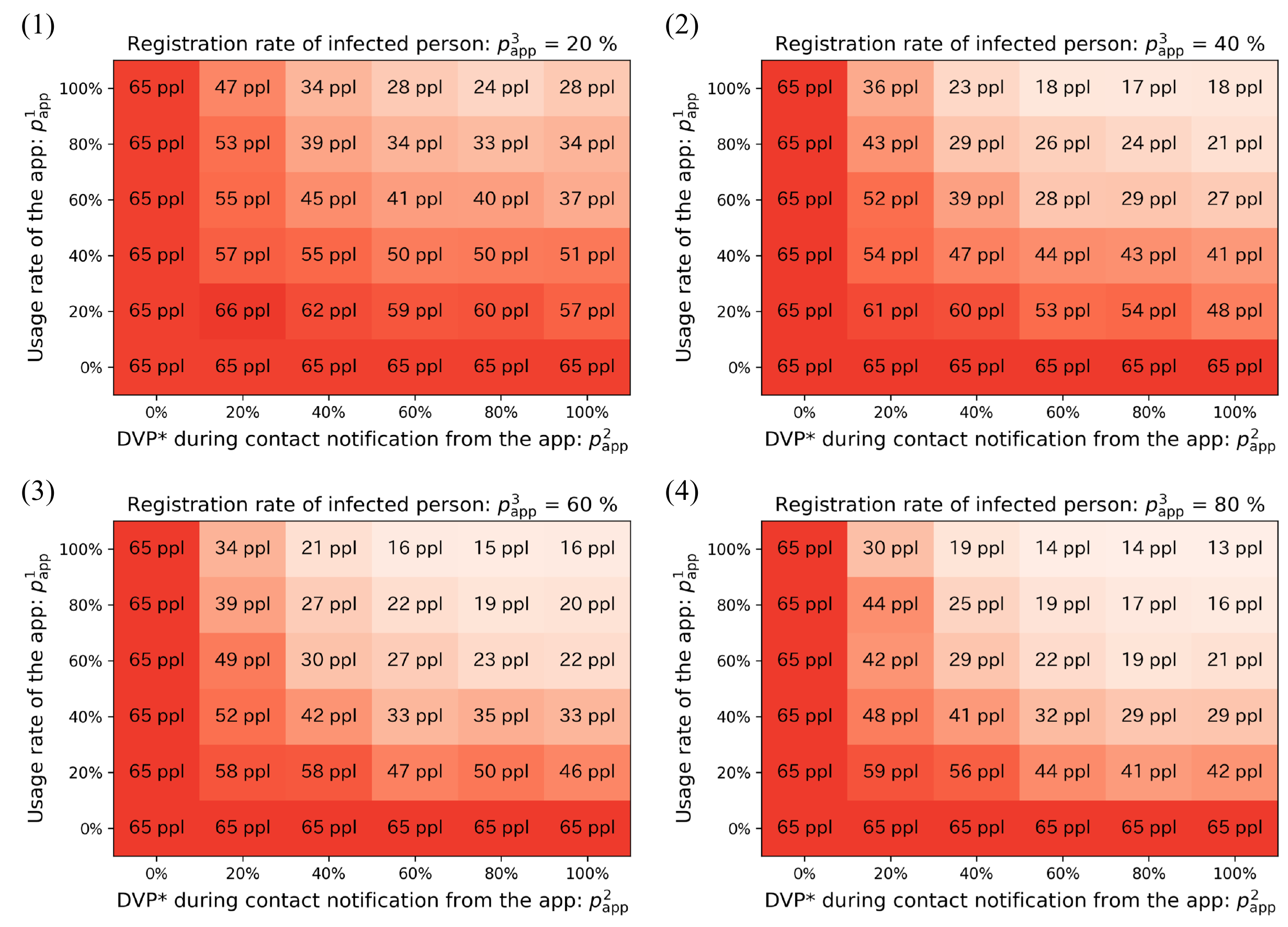}
\caption{The relationship between app parameters $p_\mathrm{app}^{\{1, 2, 3\}}$ and the number total of infectors [ppl] at the end of the 45 days simulations (in the case of $p_\mathrm{app}^{3}=20, \cdots, 80\%$). The higher the number of infectors, the redder. DVP*: Decreasing value of going out probability.}
\label{fig:4map}
\vspace{-20pt} 
\end{figure*}

Moreover, the above results are the scenarios in which all infectors who use the app registered that they are infected ($p_\mathrm{app}^{3} = 100\%$).
COCOA does not leak infection information, but infectors may be afraid of their infection information being leaked.
As a result of this, it is possible that infectors feel unwilling to register when they are infected.
To verify this scenario, we changed $p_\mathrm{app}^{3}$ from 20\% to 80\% in an increment of 20\%.
The results obtained by this procedure are shown in {\bf Fig.\ref{fig:4map}}.
Besides, we note that {\bf Figs.\ref{fig:4map} (1), (2), (3), and (4)} are the results of $p_\mathrm{app}^{3}=20\%, 40\%, 60\%$, and $80\%$, respectively.
Moreover, as $p_\mathrm{app}^{3}$ increases, the total number of infectors decreases. 
In the case of infection registration rate $p_\mathrm{app}^{3}=20\%$, to reduce the total number of infectors to half, the usage rate of the app $p_\mathrm{app}^{1}=100\%$ is required. 
This is a very difficult condition.
Moreover, in the case of $p_\mathrm{app}^{3}=40\%, 60\%$, if the usage rate of app $p_\mathrm{app}^{1}$ is at least 60\% or more, then the scenarios that reduce the total number of infectors to half appears.
Thus, because rapidly increasing the app's usage rate is difficult, it is important to register infection information.

\vspace{-10pt} 
\subsection{Results of discussions: the relationship between app parameter $\bm{p}_\mathrm{app}$ and time series trends of the total number of infectors}
In Subsection \ref{sub32}, we discussed the relationship between the app parameters $\bm{p}_\mathrm{app}$ and the number of infectors at the end of the 45 days simulation.
This discussion is important, but it is unclear whether or not the number of infectors increases after the max simulation period.
To clarify this viewpoint, it is necessary to check the time series data of the number of infectors in each scenario.
However, considering that we simulated many scenarios, it is difficult to show all the time series data.
Therefore, we calculate the index to determine whether the number of infectors has exponential, linear, or logarithmic growth.
Afterward, we discuss how to converge infection spread based on this index.

\begin{figure}[tb]
\centering
\includegraphics[scale=0.27]{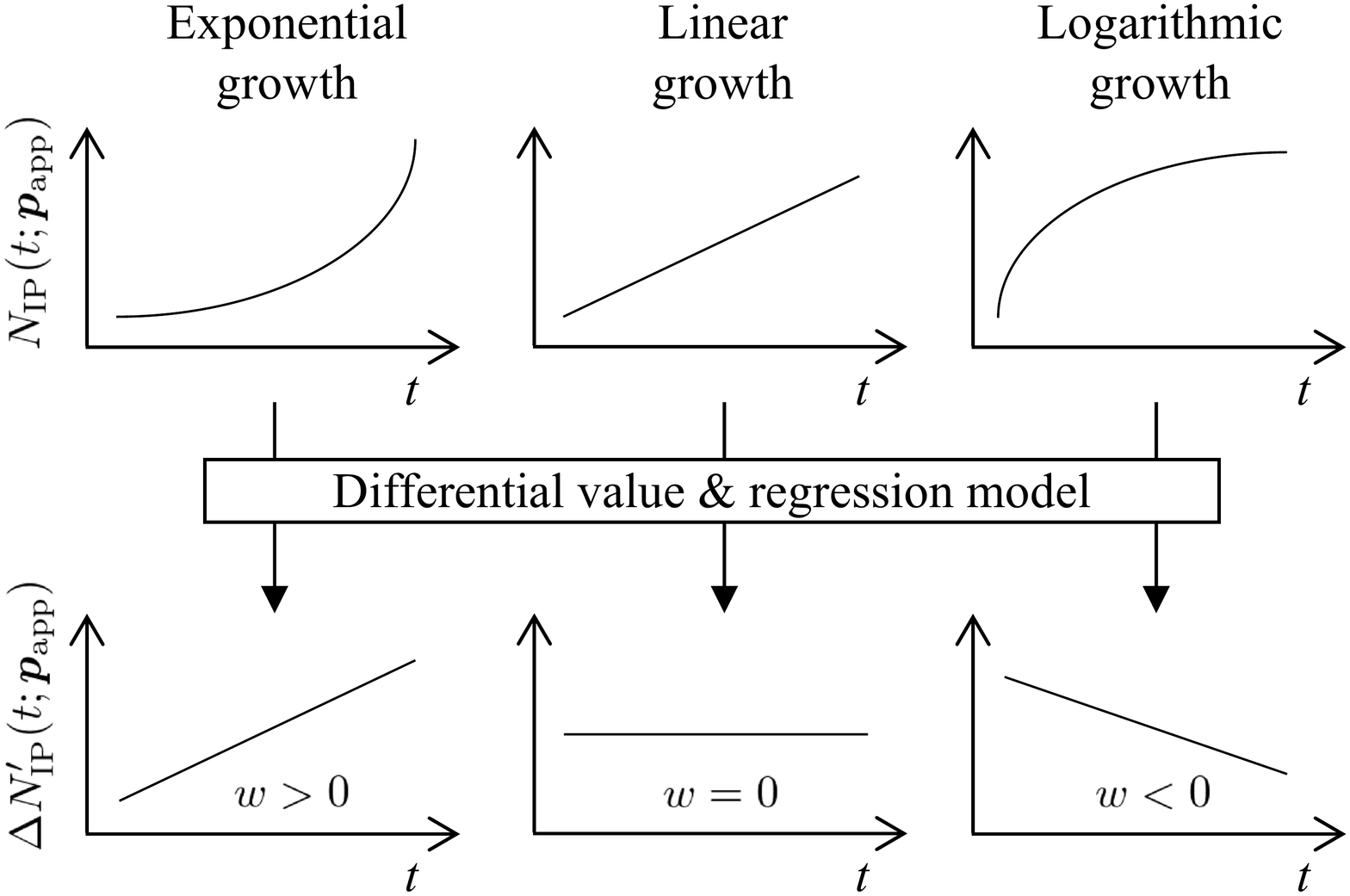}
\vspace{-10pt} 
\caption{The relationship between coefficient $w$ and the total number of infectors $N_{\mathrm{IP}}(t; \bm{p}_\mathrm{app})$}
\label{fig:ex}
\end{figure}

Now, the amount of the difference of the total number of infectors defined by
\begin{eqnarray}
\Delta N_{\mathrm{IP}}(t; \bm{p}_\mathrm{app}) = N_{\mathrm{IP}}(t; \bm{p}_\mathrm{app}) - N_{\mathrm{IP}}(t-1; \bm{p}_\mathrm{app}) \label{eq:diff}
\end{eqnarray}
is calculated,
where $N_{\mathrm{IP}}(t; \bm{p}_\mathrm{app})$ denotes the total number of infectors in $t$ days of scenario using the app parameter $\bm{p}_\mathrm{app}$, and $\Delta N_{\mathrm{IP}}(t; \bm{p}_\mathrm{app})$ denotes the differential value.
Moreover, we develop a linear regression model with the following intercept: 
\begin{eqnarray}
\Delta N'_{\mathrm{IP}}(t; \bm{p}_\mathrm{app}) = wt + b, \label{eq:w}
\end{eqnarray}
where 
\small
\begin{equation}
(w, b) = 
\mathrm{Argmin}_{w', b'} \sum_{t \in T_\mathrm{max}} ( (w't + b') - \Delta N_{\mathrm{IP}}(t; \bm{p}_\mathrm{app}))^2.
\end{equation}
\normalsize
Here, $\Delta N'_{\mathrm{IP}}(t; \bm{p}_\mathrm{app})$ denotes an estimated value of $\Delta N_{\mathrm{IP}}(t; \bm{p}_\mathrm{app})$, and $T_\mathrm{max}$ denotes the max simulation period ($T_\mathrm{max}=45$ days).
Then, the coefficient of linear regression model $w$ is the index for expressing time series trend for the total number of infectors $N_{\mathrm{IP}}(t; \bm{p}_\mathrm{app})$.
The meaning of the index is shown in {\bf Fig.\ref{fig:ex}}.
The upper side in {\bf Fig.\ref{fig:ex}} demonstrates diagram of $N_{\mathrm{IP}}(t; \bm{p}_\mathrm{app})$.
The left and center or right sides of {\bf Fig.\ref{fig:ex}} demonstrate exponential and linear or logarithmic growth, respectively.
The underside in {\bf Fig.\ref{fig:ex}} represents regression value $\Delta N'_{\mathrm{IP}}(t; \bm{p}_\mathrm{app})$ calculated by utilizing Eq.(\ref{eq:w}).
As depicted in {\bf Fig.\ref{fig:ex}}, $w > 0$, $w = 0$ or $w < 0$ represents exponential, linear or logarithmic growth, respectively. 
Therefore, since the exponential growth of the number of infectors ($w>0$) causes infection pandemic, it should be avoided.
However, because the logarithmic growth ($w<0$) leads to the convergence of infection, it is a desirable case.

\begin{figure}[t]
\centering
\includegraphics[scale=0.7]{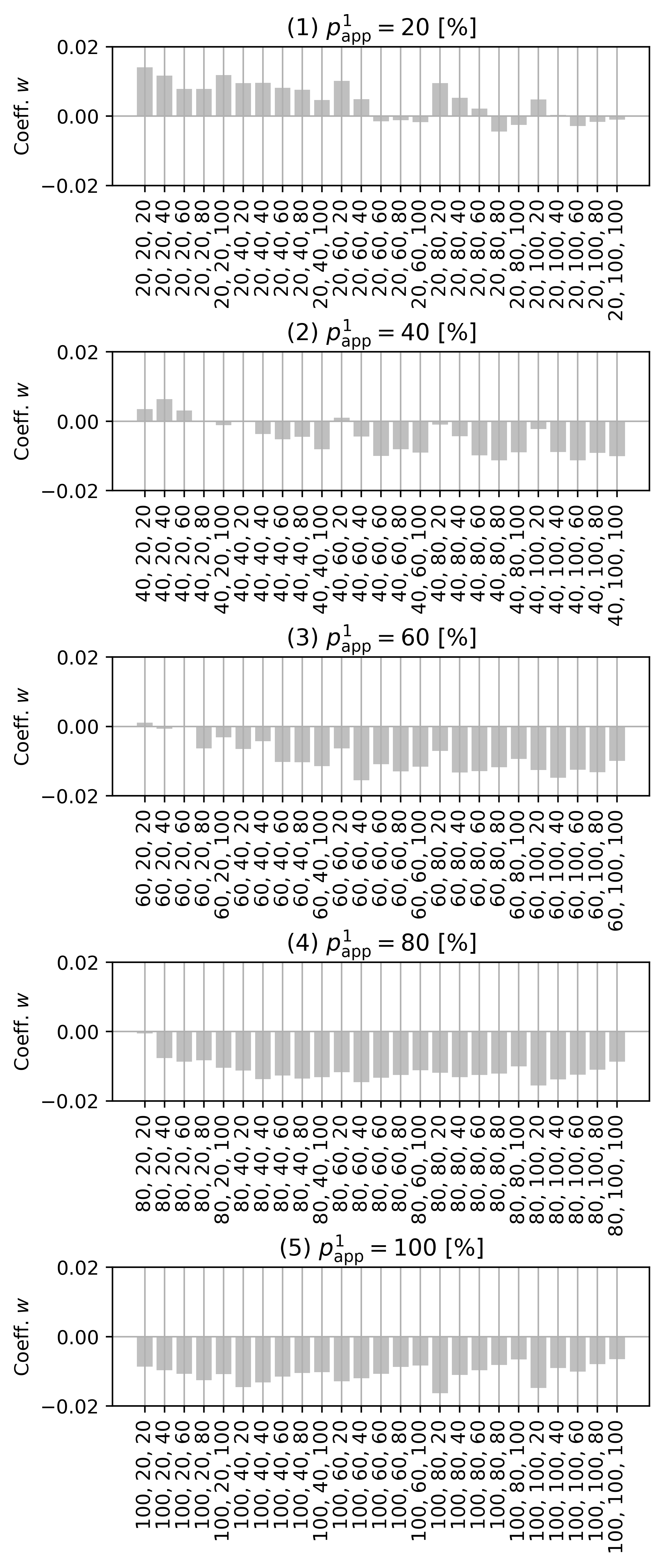}
\caption{Coefficient $w$ of each scenario. The three numbers in the horizontal axis are app parameters $p_\mathrm{app}^{\{1, 2, 3\}}$ from bottom to up.}
\label{fig:all}
\end{figure}

We calculate coefficient $w$ of all the scenarios using the app parameters $\bm{p}_\mathrm{app}$.
The results are shown in {\bf Fig.\ref{fig:all}}.
The vertical and horizontal axes represent $w$ and scenarios, respectively.
The three numbers of the horizontal axis are app parameters $p_\mathrm{app}^{\{1, 2, 3\}}$ from bottom to up.
{\bf Figs.\ref{fig:all}(1), (2), (3), (4), and (5)} denote coefficient $w$ of the usage rate of app $p_\mathrm{app}^{1}=20\%$, $40\%$, $60\%$, $80\%$, and $100\%$, respectively.

First, we discuss the case of $p_\mathrm{app}^{1} = 20\%$ shown in {\bf Fig.\ref{fig:all} (1)}.
In this case, if $p_\mathrm{app}^{1} = 20\%$ and $p_\mathrm{app}^{2}$ is 40\% or less, the signs of coefficient $w$ are positive. It illustrates an exponential growth.
In contrast, if $p_\mathrm{app}^{1} = 20\%$ and $p_\mathrm{app}^{2}$ is 60\% or more, coefficient $w$ is usually approximately 0. 
Therefore, to avoid infection pandemic during the periods of low app usage rate, it is important to keep $p_\mathrm{app}^{2}$ at 60\% or more.

Second, we discuss the case of $p_\mathrm{app}^{1} = 40\%$ shown in {\bf Fig.\ref{fig:all} (2)}.
In this case, the exponential, linear, and logarithmic growth are mixed.
In the case of $p_\mathrm{app}^{1} = 40\%$ and $p_\mathrm{app}^{2}=20\%$, there is an exponential or linear growth.
In addition, if $p_\mathrm{app}^{2}$ is 40\% or more, the growth of infectors is almost logarithmic.
Thus, when the usage rate of app $p_\mathrm{app}^{1}$ and decreasing value of going out $p_\mathrm{app}^{2}$ are 40\% or more, scenarios of convergence of infection appear. 
However, if $p_\mathrm{app}^{1}$ and $p_\mathrm{app}^{2}$ are 40\% or more and $p_\mathrm{app}^{3}$ is 20\%, then coefficient $w$ is approximately 0.
In this case, the number of infectors keeps increasing in proportion to time.
Therefore, it is important to call on infected persons to register and increase $p_\mathrm{app}^{3}$.

Finally, we discuss the case of $p_\mathrm{app}^{1}=60\%, 80\%$, and $100\%$ shown in {\bf Fig.\ref{fig:all} (3)--(5)}.
In this case, the growth of the total number of infectors is almost logarithmic.
Therefore, if about 60\% or more of all population uses the app, then the infection pandemic may converge.
As of 2019 in Japan, the rate of spread of smartphone is 83.4\%\cite{sumaho}.
Therefore, if about 70\% of smartphone users do not use the app, it is difficult to reach the usage rate of the app, which is about 60\% of all the population.
It is difficult to achieve this condition.
Thus, knowing the condition for overcoming the spread of COVID-19 is important.

Besides, the result of overcoming COVID-19 by applying the usage rate of the app, which is 60\% or more, is similar to those of Hinch et al.\cite{tr1} and Kurita et al.\cite{cocoa_kurita}.
To overcome the spread of COVID-19, Hinch et al.\cite{tr1} reported that the usage rate of the app, i.e., about 56\% of the target populations is required.
Table 1 of Kurita et al.\cite{cocoa_kurita} shows that the number of COVID-19 reproductions is less than 1.0 in many cases when the usage rate of the app is about 50\% or more. This means that the spread of COVID-19 converges.
Therefore, our study supports the results of Hinch \cite{tr1} and Kurita \cite{cocoa_kurita}.
It is noteworthy that similar results were obtained using various methods for verifying the effectiveness of the app, such as COCOA.

\vspace{-10pt} 
\section{Conclusion}
In this study, we utilized the MAS to verify the effectiveness of an app, such as COCOA.
As basic trends, as app parameter $\bm{p}_\mathrm{app}$ increases, the total number of infectors decreases (see {\bf Figs.\ref{fig:1map} and \ref{fig:4map}}).  
Therefore, the usage rate of the app, decreasing value of the going out probability of persons who had contact with infectors, and the registration rate of infectors can effectively reduce the spread of COVID-19.
Moreover, if the usage rate of the app is 60\% or more, the time series trends of the number of infectors in many scenarios have logarithmic growth (see {\bf Fig.\ref{fig:all}}).
Our study's result supports some previously conducted studies \cite{tr1, cocoa_kurita}.
In the case of the simulation task that cannot verify reliability in the real world, the accumulation and comparison of case studies using various simulation methods is important.
Thus, this study's result is beneficial.

As of Aug. 7, 2020, the installation rate of the COCOA of all the Japanese population is about 10\%\cite{ref:6p}.
Besides, since one is required to always turn on his/her smartphone's bluetooth, the actual usage rate of the app will be lower than the installation rate.
As of Aug. 2020 in Japan, it is very important to call on everyone to install COCOA.
We observed that one of the reasons that many peoples do not install COCOA is because they do not know its effectiveness in the reduction of infectors.
To solve this problem, it is very important that researchers report the effectiveness of this app even if the simulation environment is performed in an artificial society.
Moreover, we hope that the accumulation of case studies, such as our study, can lead to an increase in the usage rate of the app.

Besides, this study reported the effectiveness of only the app.
As mentioned in Section \ref{sec:pre}, our simulator can include the capacity limit of isolation wards.
In our future studies, we will report the effectiveness of multiple measures, such as mixing the app and capacity limit of isolation wards to overcome the spread of COVID-19.
Finally, we will announce the effectiveness of the app and other measures to everyone and request them to install and use the App.
We believe that apply consistent efforts can help us to overcome the spread of COVID-19.



\end{document}